\documentclass[fleqn,10pt]{article}
\usepackage{fullpage}
\usepackage{graphicx,hyperref}
\usepackage{amsmath,amssymb}
\usepackage{color}
\usepackage{times}
\usepackage[noend]{algpseudocode}

\usepackage{fullpage}
\usepackage{amsfonts}
\usepackage{amsmath}
\usepackage{multirow}
\usepackage{longtable}
\usepackage{graphicx}
\usepackage{sidecap}
\usepackage{subcaption}
\usepackage[affil-it]{authblk}
\usepackage{hyperref}
\usepackage{cite}
\usepackage{color}

\begin{document}

\title{\textbf{Mapping temporal-network percolation to weighted, static event graphs}}
\date{\vspace{-5ex}}

\author[1]{Mikko Kivel\"a}
\affil[1]{\normalsize{Department of Computer Science, Aalto University School of Science, FI-00076 AALTO, Finland}}

\author[2,1]{Jordan Cambe}
\affil[2]{\normalsize{Univ Lyon, ENS de Lyon, Inria, CNRS, UCB Lyon 1, LIP UMR 5668, IXXI, F-69342, Lyon, France}}

\author[1]{Jari Saram\"aki}

\author[2]{M\'{a}rton Karsai \thanks{Corresponding author: marton.karsai@ens-lyon.fr}}

\maketitle

\begin{abstract}
Many processes of spreading and diffusion take place on temporal networks, and their outcomes are influenced by correlations in the times of contact. These correlations have a particularly strong influence on processes where the spreading agent has a limited lifetime at nodes: disease spreading (recovery time), diffusion of rumors (lifetime of information), and passenger routing (maximum acceptable time between transfers). Here, we introduce weighted event graphs as a powerful and fast framework for studying connectivity determined by time-respecting paths where the allowed waiting times between contacts have an upper limit. We study percolation on the weighted event graphs and in the underlying temporal networks, with simulated and real-world networks. We show that this type of temporal-network percolation is analogous to directed percolation, and that it can be characterized by multiple order parameters.
\end{abstract}

\section*{Introduction}



Contact network structure plays an important role in many dynamical processes, in particular in diffusion and spreading~\cite{barrat2008dynamical,pastorsatorras2015}. Recently, it has been realized that also the temporal properties of networks have major effects on the dynamics of spreading~\cite{saramaki_holme_2012,Holme2015,masudalambiotte2016}. This is (a) because spreading processes have to follow causal, time-respecting paths spanned by sequences of contacts, \cite{Kempe2002,Moody2002,holme2005network,pan2011path,scholtes2014causality} and (b) because the speed of spreading and the ability of spreading processes to percolate through the contact structure are heavily influenced by temporal inhomogeneities \cite{Iribarren2009,Karsai2011,kivela2012multiscale,Horvath2014,delvenne2015} and correlated contact times 
\cite{pfitzner2013,Backlund2014,Aoki2016}. However, this picture is still lacking detail, especially when it comes to percolation.

Processes with limited waiting times at nodes are particularly sensitive to temporal inhomogeneities. These include variants of common disease spreading models such as the Susceptible-Infectious-Recovered (SIR) and Susceptible-Infectious-Susceptible (SIS) \cite{Iribarren2009,Miritello2011,rocha2011simulated,Lee2012,holme2014birth,holmemasuda2015,valdano2015,Genois2015,HolmePRE2016,Speidel2016}, where nodes only remain infectious for finite periods of time. Other examples include social contagion \cite{DaleyKendall,socdynrev}, ad-hoc message passing by mobile agents \cite{TrippBarba20161}, and passenger routing in transport networks \cite{Nassir201626}.  In all these processes, the spreading agent must be transmitted onwards from a node within some time $\delta t$, or the process stops.

This maximal allowed waiting time $\delta t$ can be incorporated into time-respecting paths by requiring that their successive contact events are separated by no more than $\delta t$ units of time. 
The existence of such time-respecting paths then determines the outcome of the spreading process. For very low values of $\delta t$, network-wide connectivity is unlikely to exist and spreading processes are unlikely to percolate the network. On the other hand, large $\delta t$ may provide the spreading process with enough pathways to infect a substantial fraction of the network.

The waiting time limit $\delta t$ is thus the control parameter of a percolation problem, where connectivity is determined by paths of contact events that follow one another within $\delta t$. However, discovering all such paths independently for each value of $\delta t$ is computationally expensive. Therefore a fast way of computing connectivity for different values of $\delta t$ is required. In this Letter, we introduce the weighted event graph as fast solution to the computational problem of temporal-network percolation, and use this representation to study percolation in artificial and real networks. We show that in temporal-network percolation there are, in fact, three types of order parameters, measured in terms of component nodes, events, and lifetime, and that temporal-network percolation has strong connections to directed percolation.

Weighted event graphs are static, weighted, and directed acyclic graphs (DAGs) that encapsulate the complete set of $\delta t$-constrained time-respecting paths for all values of $\delta t$ simultaneously. The subset of paths corresponding to a specific value of $\delta t$ can be quickly extracted from the weighted event graph by a simple thresholding procedure.
Weighted event graphs can be viewed as a temporal-network extension of the line-graph representation of static networks, while they also share some similarity with the approach of Ref.~\cite{scholtes2014causality} that maps two-event sequences onto aggregated second-order networks, as well as that of Ref.~\cite{Mellor2017The} where an unweighted event graph is constructed from pairs of temporally closest events. The approach presented in this Letter builds on concepts introduced in Refs.~\cite{backlund2014Thesis,ayala2015}.


\begin{figure}
\centering
\includegraphics[width=.6\linewidth]{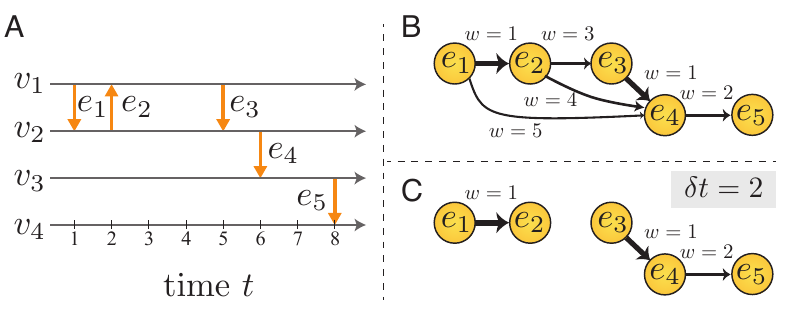}
\caption{Constructing and thresholding the weighted event graph. a) A timeline representation of a temporal network with four nodes $v_1-v_4$ and five events $e_1-e_5$. b) The weighted event graph representation of the temporal network. c) The thresholded event graph, containing only pairs of events with a maximum time difference of $\delta t=2$.}
\label{Fig:1}
\end{figure}

Let us consider a temporal network $G=(V,E,T)$ with edges defined as a set of events $E \subset V \times V \times [0,T]$ over a time period $T$. For simplicity, we allow no self-edges 
or simultaneous events of the same node. 
Two events $e=(u,v,t)$ and $e'=(u',v',t')$ are considered adjacent so that $e\rightarrow e'$ if they share at least one node and $t< t'$. This definition of the adjacency is directed and preserves the arrow of time. Further, two adjacent events are considered $\delta t$-adjacent if their time difference is $0<t'-t<\delta t$.  The weighted event graph representation of a temporal network $G$ is defined as the graph $D=(E,E_{D},w)$ where the set of nodes $E$ is the set of events in $G$ and the edges in $e_D \in E_{D}$ represent the adjacency of the events $e_{D}=e\rightarrow e'$ with weights defined as $w(e_{D})=t'-t$. Note that $D$ contains all time-respecting paths in the network. For paths where the longest allowed waiting time is $\delta t$, we can get the subgraph $D_{\delta t}$ by thresholding $D$ so that only links with $w\leq\delta t$ are retained. This allows us to sweep through the whole range of $\delta t$  with minimal computational cost (for details on algorithms, see Supplementary Informations (SI)). 

The $\delta t$-thresholded event graph $D_{\delta t}$ is a superposition of the time-respecting paths that a $\delta t$-limited spreading process may follow. Therefore, its structure tells if the process can percolate the network. A closer look at the problem reveals that here, the concept of percolation is more complex than for static networks.
Let us first look at the component structure of $D_{\delta t}$. It is directed, but may only contain weakly connected components; there are no strongly connected components because $D_{\delta t}$ is by definition acyclic. Each event graph node can be associated with an in-component and out-component that contain the events on up- and downstream temporal paths; note that these components naturally overlap for different event graph nodes. In the following, we will limit our analysis to weakly connected components of $\delta t$ because of their uniqueness, unless stated otherwise; note that for spreading processes, the existence of a weakly connected component is a necessary but not a sufficient condition for percolation. 

Let us next address the question of connected component size. In percolation analysis, the relative size of the largest  connected component is defined as the order parameter, while the quantity equivalent to magnetic susceptibility is often identified as the average size of the other connected components. Here, there are three ways of measuring the size of a component of $D_{\delta t}$. (1) The most straightforward way is to count the  number of \textit{event graph nodes} $S_{E}(E')=|E'|$ in a connected component $E' \subseteq E$ of $D_{\delta t}$. This is the same as the maximal of events on the component's time-respecting paths that any spreading process can follow. (2) One can count the  number of \textit{nodes of the temporal network} $S_{G}(E')=|\bigcup_{(u,v,t)\in E'}(u \cup v)|$ that are covered by the event graph component $E'$. This is an upper bound for the number of temporal network nodes that any spreading process can reach by following the component's time-respecting paths. (3) One can measure the \textit{lifetime} of the event graph component $S_{LT}(E')=(\max_{(u,v,t)\in E'} t-\min_{(u,v,t)\in E'} t)$. This is the maximum possible lifetime of any spreading process on the component. Note that there may be several co-existing components with long (or even infinite) lifetimes; frequent and sustained contacts between a small number of nodes can already give rise to a long-lived component.

With the above measures, we can define the order parameter and susceptibility as 
\begin{eqnarray}
\rho_{*}(D_{\delta t}) &=& \dfrac{1}{N_{*}} \max_{E' \in D_{\delta t}} S_*, \label{eq:rho} \\
\chi_{*}(D_{\delta t}) &=& \dfrac{1}{N_{*}} \sum_{{S_*}<\rho_{*}} n_S S_*^2,  \label{eq:chi}
\end{eqnarray}
where $n_S$ is the number of components of size $S_*$, and $N_{*}=\sum_{S_*} n_S S_*$ for the chosen definition of size $* \in \{E,G,LT\}$.

Note that the above picture has a clear link to \emph{directed percolation} ~\cite{Hinrichsen2000Non}, where there are two correlation lengths, temporal and spatial, characterizing correlations parallel and perpendicular to the directed lattice.
In our case, the arrow of time provides the direction. However, instead of the regular lattice of the usual directed-percolation picture, our process unfolds on a highly irregular structure determined by the set of events that take place at each moment in time. In this setting $\rho_{E}$ gives the probability that a randomly selected event in $D_{\delta t}$ belongs to a structurally percolating infinite cluster, while $\rho_{LT}$ is the typical temporal correlation length for a given $\delta t$. Note that in our case these correspond to two different order parameter definitions as the largest and longest components might not be the same, as they typically are in directed percolation.

To explore how $\delta t$ controls connectivity in  temporal networks, we introduce a simple toy model. We define an ensemble of temporal networks $\mathcal{G}_{p,r}(n,k,\alpha)$ where the topology is that of an Erd\H{o}s-R\'enyi (E-R) random graph with $n$ nodes and average degree $k$, and events are generated on each link by a Poisson process with $\alpha$ events per link on average.  We set the observation period $T$ long enough so that  $\delta t \ll T$ and $\alpha \ll T$.

In this model, there is a transition from the disconnected to the connected phase when the independent 
Poisson events become $\delta t$-adjacent and form a giant 
weakly connected component in $D_{\delta t}$. In terms of degree, a lower bound for this critical point can be estimated as the point where the average out-degree of the event graph becomes $\langle k^{out}_{_{D_{\delta t}}}\rangle=1$. In the underlying E-R network, each edge is adjacent to  $2(k-1)+1$ edges (including the edge itself), and therefore
the average out-degree of $D_{\delta t}$ is $\langle k^{out}_{D_{\delta t}} \rangle=\alpha \delta t \left[2(k-1)+1\right]$. The condition for the critical point  can then be written as
\begin{equation}
k_c= \frac{(\alpha \delta t)^{-1} -1}{2}+1  \hspace {.2in}\mbox{and}\hspace{.2in} \delta t_c= \frac{1}{\alpha(2k-1)}.
\label{eq:line}
\end{equation}

\begin{figure}
  \centering
\includegraphics[width=.6\linewidth]{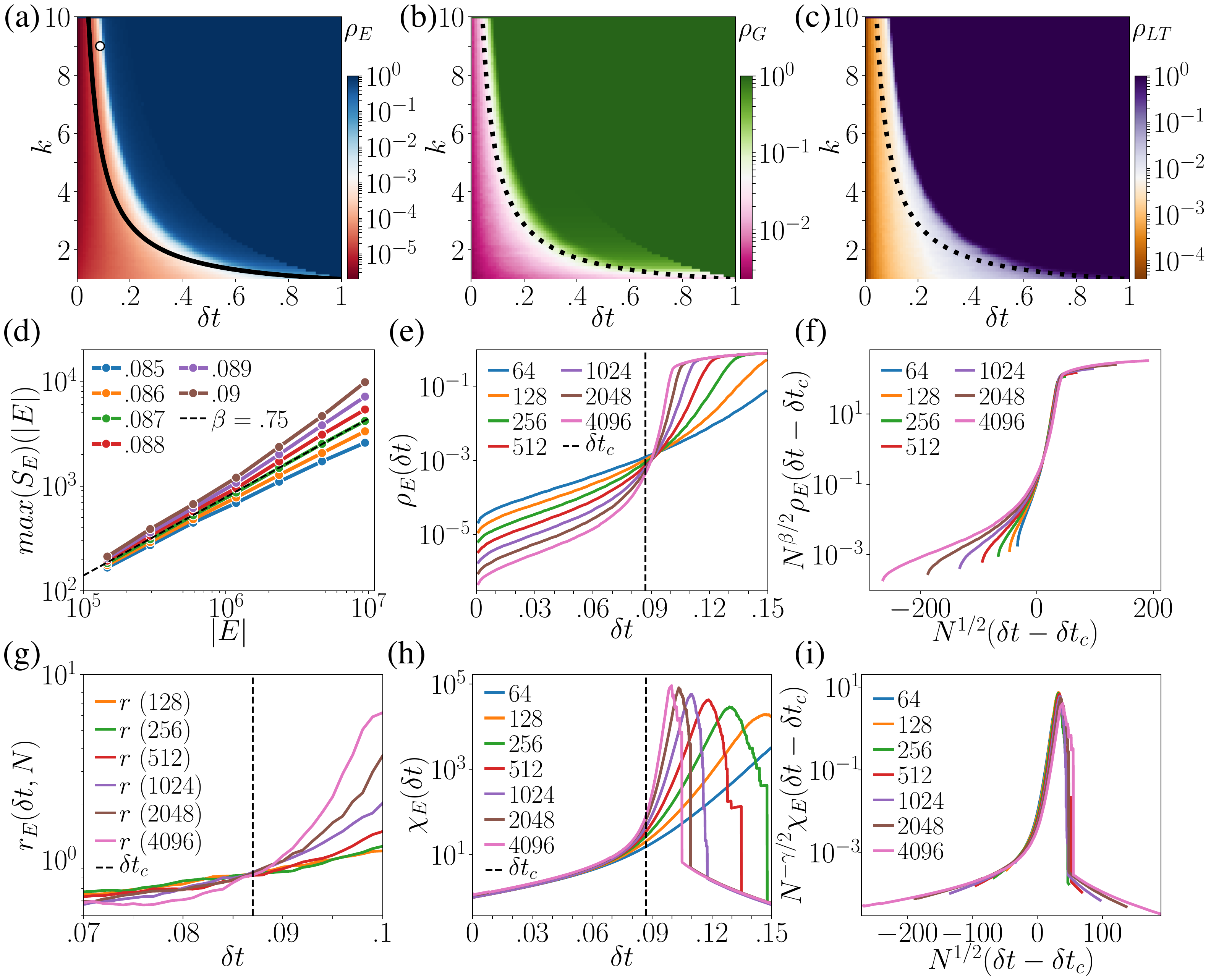}
\caption{Phase diagrams for the random temporal network model as a function of the average degree $k$ of the underlying topology and the maximal allowed time difference between events, $\delta t$. The colormaps show the (ensemble-averaged) fractional size $\rho_{*}(k,\delta t)$ of the giant weakly connected components, measured in terms of (a) number of events in the event graph components $S_E$, (b) number of nodes of the temporal network $G$ that the $S_E$ largest event graph component covers, and (c) the lifetime of the event graph component $S_{LT}$. The solid line in (a) depicts the analytic estimate of Eq.\ref{eq:line}, while the circle in the upper left corner shows the critical point for $k=9$ determined as explained in the text. The dashed lines in (b) and (c) show the estimate of Eq.\ref{eq:line} computed for $S_E$ for reference. (d) Scaling of $\max(S_E)$, the size of the largest weakly connected component in $D_{\delta t}$, with the size of $D_{\delta t}$ measured in number of event-nodes $|D_{\delta t}|=|E|$, for different $\delta t$. The dashed line corresponds to the critical $\delta t_c=0.87$. (e) The order parameter curves $\rho_E(\delta t)$ for different network sizes $N=|V|$ with the determined $\delta t_c$ shown as a dashed line. (f) Same as (e) after finite-size scaling using the function defined in Eq.\ref{eq:rhofs}. (g) The ratios $r(\delta t,N)$ crossing at $\delta t_c$. The dashed line shows the critical point determined in (d). (h) Susceptibility curves $\chi_{w}(\delta t)$ for different sizes with the determined $\delta t_c$ shown as a dashed line. (i) Same as (h) after finite-size scaling using the function defined in Eq.\ref{eq:chifs}. Computations on panels (a-c) were carried out with a model network of $|V|=2048$ nodes evolving for $T=512$ with an event rate of $\alpha=1$ averaged over $10$ realizations. Results on panels (d-i) had the same parameters but were were averaged over $100$ realizations and may differ in size.}
\label{fig:2}
\end{figure}

 
 This theoretical line $\delta t_c(k)$ is shown together with results of simulations in Fig.\ref{fig:2}.a, using the number of events as the measure for the fractional size of the largest component  $S_E$. $\delta t_c(k)$ is seen to separate the percolating and non-percolating regimes in the simulation fairly well. Fig.\ref{fig:2}.b and c show the relative largest component sizes (i.e. the order parameter) measured in terms of temporal-network nodes ($S_G$) and component lifetime ($S_{LT}$); there is indication of percolation transition happening close to the theoretical line $\delta t_c(k)$ computed for the number of events. Note that this cannot be expected generally; the phase transition lines for events, nodes, and lifetime can well be different. 

Let us investigate the critical behavior in the model in detail, fixing the average degree to $k=9$. This makes the thresholded event graph $D_{\delta t}$ dense enough to motivate a mean-field (MF) approach; in case of regular lattices MF solution gives a good approximation above the critical dimension $d_c=4$. We locate the critical point with two independent methods. First, we exploit that once the system reaches a stationary state where the order parameter becomes time-invariant beyond statistical fluctuations, a scaling relation $max(S_E) \sim |D_{\delta t}|^{\beta}$ is expected to hold around the critical point $\delta t_c$, where $|D_{\delta t}|$ is the size of the thresholded event graph measured in events (event-graph nodes), and $\beta$ is the critical exponent of the order parameter. We measured this relation for several values of $\delta t$ and several system sizes (see Fig.\ref{fig:2}d) and found a power-law scaling of $S_E(|D_{\delta t}|)$ around the critical point $\delta t_c\simeq 0.087$ with the exponent $\beta\simeq 0.75$.  This point is shown as a circle in Fig.\ref{fig:2}a; it is above the analytical estimate which provides a lower bound for the critical point. Note that for the directed-percolation university class, the MF solution suggests $\beta_{MF}=1$.

The second, independent way of determining the critical point is to calculate the ratios $r(\delta t, N)=\rho_E(\delta t,N)/\rho_E(\delta t,N/2)$ for varying $N$. These should cross  around the critical point $\delta t_c$ where $r(\delta t_c,N)=2^{-x}$, where $x$ is related to the finite-size scaling exponent. In Fig.\ref{fig:2}g, this point of crossing indeed appears  close to $\delta t_c\simeq 0.087$ with a value $r(\delta t)\simeq 0.82$ suggesting an exponent $x\simeq 0.2863$, which should be compared to $\beta/2\simeq 0.375$.

Finite-size scaling in networks is naturally related to the network volume $N$ (number of nodes) instead of a linear size scale $\ell$, which cannot usually be defined. Assuming the correspondence $N \leftrightarrow \ell^d$, one can derive the finite-size scaling functions, which are expected to hold in the conventional mean-field regime, i.e. if $d>d_c$, or in other words if the network is dense enough as in our case.
This leads to a finite-size scaling function of magnetisation:
\begin{equation}
\rho_{E}(\delta t,N)\sim N^{-\beta / d \nu^*}\widetilde{\rho}_{E}(N^{1/ d \nu^*}(\delta t-\delta t_c)),
\label{eq:rhofs}
\end{equation}
where $\nu^*=2/d$ is the finite-size scaling exponent (of linear size) which depends on the dimension $d$. If $d<d_c$ it is the spatial correlation length exponent, and above the critical dimension $d_c=4$ it takes the value $\nu^*=2/d_c$~\cite{Karsai2006Nonequilibrium}. 
At the same time a similar scaling function is expected to hold for susceptibility:
\begin{equation}
\chi_{E}(\delta t,N)\sim N^{\gamma / d \nu^*}\widetilde{\chi}_{E}(N^{1/ d \nu^*}(\delta t-\delta t_c)),
\label{eq:chifs}
\end{equation}
where $\gamma$ is the mean cluster-size exponent. From the definition of $\chi_{E}$ (in Eq.\ref{eq:chi}) and the scaling of $\rho(\delta t,N)$ at $\delta t_c$ we can derive the simple exponent relation $\gamma/(d \nu^*)=1-\beta/(d \nu^*)$, where $\nu*=2/d$, $d=d_c=4$ and $\beta\simeq 0.75$, which gives us a value $\gamma\simeq 1.25$ (which is slightly different from the directed-percolation MF value of $\gamma_{MF}=1.0$).

To check whether the predicted finite-size scaling behaviour holds around the critical point, we took the simulated $\rho_{E}(\delta t,N)$ and $\chi_{E}(\delta t,N)$ measured for various $N$ (see Fig.\ref{fig:2}e and h respectively). Using the scaling functions in Eq.\ref{eq:rhofs} and Eq.\ref{eq:chifs} with the determined exponents, we scaled the order parameter and susceptibility curves as the function of $(\delta t-\delta t_c)$. As shown in Fig.\ref{fig:2}.f and i (respectively), we obtained the expected scaling behaviour for both quantities in the vicinity of the critical point.

\begin{figure}
  \centering
\includegraphics[width=.6\linewidth]{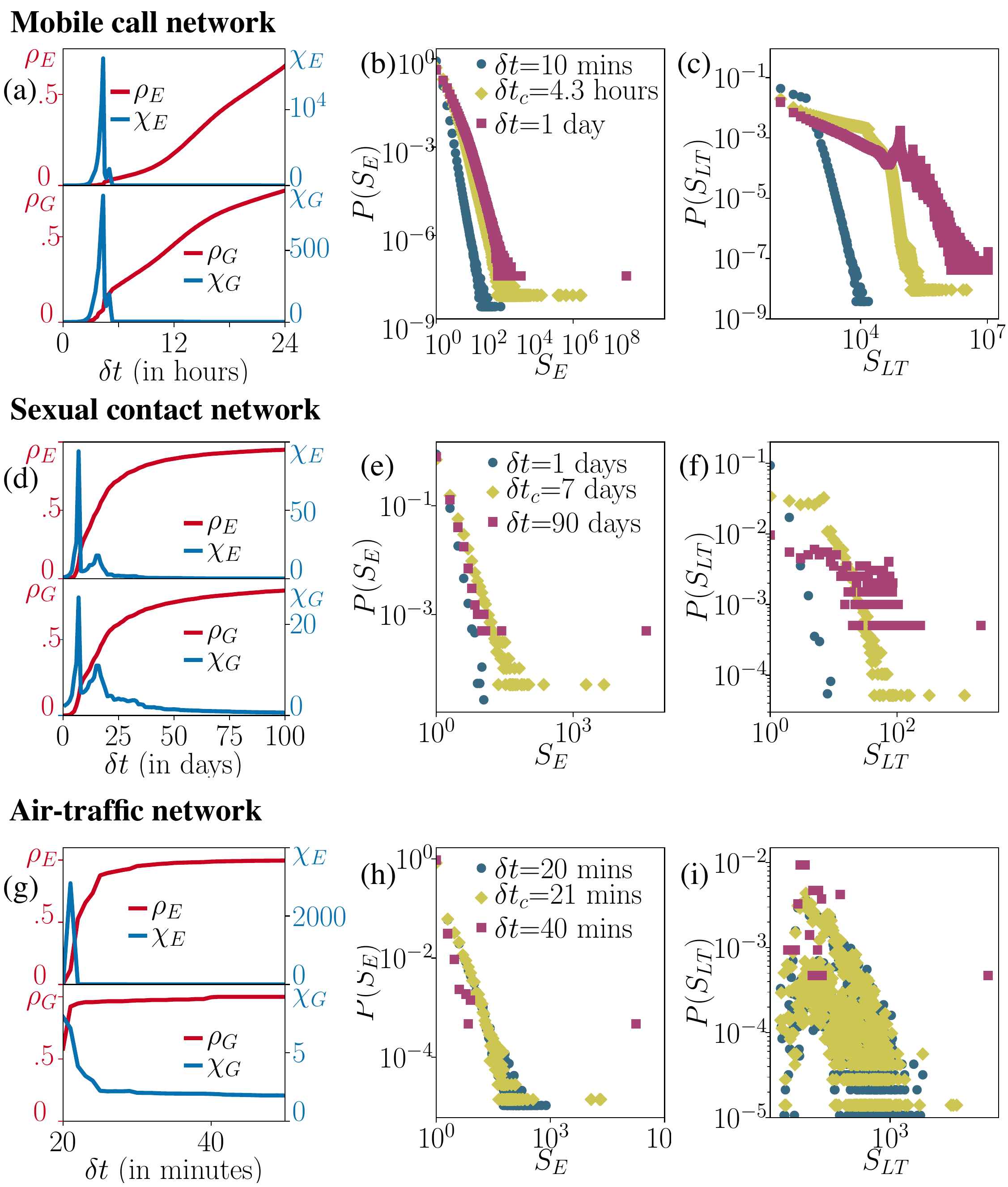}
\caption{Percolation transitions in empirical temporal networks of mobile communication (a, b, c), sexual interactions (c, d, e), and air-transportation (g, h, i). Panels (a, d, g) depicts the order parameter $\rho_*(\delta t)$  (solid red lines) and susceptibility $\chi(\delta t)$  (solid blue lines) of weakly connected components 
with sizes measured in events  (upper panels) and temporal-network nodes (lower panels). Panels (b, e, h) show the size distributions of weakly connected components in the event graph measured below (blue circles), at (yellow diamonds), and above (violet squares) the critical $\delta t_c$. Panels (c, f, i) are the same but depict the distribution of the lifetimes of weakly connected event graph components.}
\label{fig:3}
\end{figure}

We next investigated the temporal percolation in real-world temporal networks. We studied three empirical networks: (a) a mobile call network~\cite{Karsai2011} of $\sim3.2\times 10^8$ time-stamped interactions over $120$ days of $\sim 5.2\times 10^6$ of customers of an European operator; (b) a sexual-interaction network~\cite{rocha2011simulated} from Brazil with $16,726$ sex workers and clients who interacted $42,409$ times over $2,231$ days; and (c) an air-transportation network with the time, origin, destination and duration of $180,192$ flights between $279$ airports in the United States~\cite{USAirline} over $10$ days. For further details, see SI. These three networks are relevant for different diffusive processes, of information, disease, and passengers.

We measured the largest weakly connected component $\rho_E$ (resp. $\rho_G$) and  susceptibility $\chi_E$ (resp. $\chi_G$) in terms events (resp. temporal-network nodes) covered by the event graph components. As seen in Fig.\ref{fig:3}a for the mobile calls and in \ref{fig:3}d for the sexual-interaction networks, the phase transitions of both types of components take place at similar times. The percolation point $\delta t_c$ is well identifiable as the peaks of susceptibility, with $\delta t_c\sim 4$ h $20$ min for the calls and $\delta t_c\sim 7$ d for the sexual interaction network. A spreading agent has to survive at the nodes at least this long if it is to spread over the entire network. Interestingly, for both networks, the susceptibility shows a second peak (resp. $\sim 5$ hours and $\sim 16$ days), which may indicate another characteristic temporal scale of importance for the connectedness.

For the air transportation network, we set further conditions. While for the other networks we assumed instantaneous and bidirectional events, for flights we consider directed events that have a specific duration (for details, see Supplementary Informations (SI)). We also set a minimum $\delta t=20$ minutes to capture the best-case achievable transfer time. As shown in Fig.~\ref{fig:3}g this network too goes through phase transitions, first in $G$ at around $20$ min, and shortly thereafter  in  $D_{\delta t}$ at around $21$ min. Moreover, after $\delta t=45$ min the  temporal network is almost entirely connected. Note that $\delta t= 21$ minutes is close to a minimal transfer time while $45$ min can be considered as the typical transfer time between flights.

Percolation theory suggests that the structural phase transition can be also captured by measuring the component size distributions around $\delta t_c$. It is expected that below the critical point ($\delta t<\delta t_c$), only connected components with exponentially small sizes are present in the system. At the critical point $\delta t_c$, $P(S_*)$ appears with a power-law tail, and above $\delta t_c$ the distribution is dominated by a single giant  component; other components are exponentially small. This behaviour has been found for the distributions of weakly connected components measured in terms of events or nodes in all three empirical systems (see Fig~\ref{fig:3}b, e, and h). However, in terms of component lifetimes, another picture emerges due to possible non--unique giant components as discussed earlier. While for small $\delta t$, the tail of $P(S_{LT})$ appears as a power-law, for larger $\delta t$ several long-lived components may exist, spanning the entire observation period, even when involving only a microscopic fraction of network nodes and events. 


Above, we focused on weakly connected components. However, for spreading processes, these only provide an upper estimate (a worst-case scenario) of the largest set that is reachable from any event graph node. For a more precise measure, one needs the largest \emph{out-component} that is computationally rather costly to obtain. This problem sets one important future direction: to develop an algorithmic solution for identifying the largest out-component of any node in dense directed networks of large sizes.

In this Letter, we introduced a new representation of temporal networks by mapping them weighted event graphs that are static, weighted, directed, and acyclic. Weighted event graphs provide a powerful tool because they contain all $\delta t$-constrained time-respecting paths. In addition to temporal-network percolation, weighted event graphs can be used to compute centrality scores for events, links, and nodes, and to identify the complete set of $\delta t$-connected temporal motifs~\cite{Kovanen2011} in a computationally economic way. Further they open new directions for studying system-level higher-order correlations in temporal networks through the analysis of a static structure.

\subsection*{Acknowledgements}
We thank A.-L. Barab\'asi for providing the mobile call dataset. JS acknowledges funding from the Academy of Finland, project n:o 297195. MK acknowledges support from the Aalto Science Institute and the SoSweet ANR project (ANR-15-CE38-0011-01).


\newpage

\begin{center}
{\LARGE Supplemental Information for}\\[0.7cm]
{\Large \textbf{Mapping temporal-network percolation to weighted, static event graphs}}\\[0.5cm]
{\large M. Kivel\"a, J. Combe, J. Saram\"aki, M. Karsai}\\[2cm]
\end{center}

\vspace{.3in}

\section{Temporal networks with durations/delays on events}
\label{sec:Durations}

In some temporal systems it is not enough to consider the times of events, but one needs to also consider
the durations (or delays) of the events \cite{Holme2012Temporal}. In this case we define a set of events as $E \subset V \times V \times [0,T] \times [0,T]$,
where in an event $(u,v,t,t_d) \in E$ the additional member $t_d$ represents the duration or delay related to the event. 
This additional element is necessary for example in the air-traffic network studied here, where the $t_d$ represents the flight
times, and allows us to consider $\delta t$-adjacent time-respecting paths that can correspond to actual trips taken
in the system. 

The $\delta t$-adjacency in systems with duration or delay are defined exactly as in the simpler systems but with the 
allowed time difference between two events $e=(u,v,t,t_d)$ and $e^\prime=(u^\prime,v^\prime,t^\prime,t_d^\prime)$ being defined
as $0 < t^\prime - t  - t_d < \delta t$. Note that this is not a separate definition for the $\delta t$-adjacency but
a generalisation of the case without durations, as the $\delta t$-adjacency defined in the main text is returned when
all events have $t_d=0$.

\section{Definitions of $\delta t$ adjacencies for directed networks}
\label{sec:DAGadj}

In directed networks spreading, diffusion, and progress of other dynamics are constrained by the direction of the
edges in addition to the arrow of time. This can be taken into account in the $\delta t$-adjacencies by restricting
the adjacencies where $e \rightarrow e^\prime$ only when in the two events $e=(u,v,t,t_d)$ and $e^\prime=(u^\prime,v^\prime,t^\prime,t_d^\prime)$
have $v=u^\prime$. The air-traffic network studied in the main text is considered directed in this way.

\section{Algorithm to construct directed acyclic graph representation of temporal networks}
\label{sec:DAGalg}

Constructing the weighted DAG representation of a temporal network $D=(E, E_D, w)$ can be done efficiently by noting that the
edges in $E_D$ can be listed by inspecting the sequence of events around each node $v \in V$ separately. 
For some data sets the full DAG $D$ might be large, and it is convenient to construct $D_{\delta t_{\max}}$ that can, for example, 
be later used to sweep through all values $\delta t < \delta t_{\max}$.

For each node in the temporal network $v \in V$ one can build a time-ordered sequence of events $\{ e_1, \dots, e_k \}$ in which $v$ participates.
In the case where there are no durations one can then simply iterate over each event $e_i$, and for each of them search forward in the ordered
event sequence until one finds an event $e_j$ for which $t_j - t_i > \delta t_{\max}$. One then adds a link $e_i \rightarrow e_j$ at the each
step of this process until the event $e_j$ that is too far from the starting event $e_i$ is found. (Note that some $\delta t$ adjacencies are found
twice.) Creating the event sequences and sorting them can be done in $\mathcal{O}(m \log m)$ time, and as each step of the algorithm produces a
single link (with possibility of some links being visited twice) the algortihm runs in total $\mathcal{O}(m \log m + |E_D|)$ time.
Including the durations of events only requires a small adjustment to this algorithm, for example, a construction of sequences of events that are
sorted according to the end times of the events $t+t_d$.

\begin{algorithmic}[1]
\Function{DAG edges for a node}{$\{ e_1, \dots, e_k \}$}
\For{$i \gets 1 \textrm{ to } k$}
 \State{$j \gets i$}
 \While{$t_j - t_i \leq \delta t_{\max}$ and $j \leq k$}
   \State{Output: $e_i \rightarrow e_j$}
 \EndWhile
\EndFor
\EndFunction
\end{algorithmic}


\section{Extracting component distributions of directed acyclic graphs}
\label{sec:DAGcompdistr}

The DAG representation turns problems related to dynamics to problems of graph structure, and this allows
one to take advantage of computational methods developed for analysing massive graphs. 

The $\delta t$ adjacencies within a specific range $\delta t < \delta t_{\max} $ can
be calculated via ``thresholding'' the full network (i.e., removing all edges below the threshold level $\delta t_{\max}$). 
In static weighed networks the weakly-connected 
component distributions can be calculated for all possible threshold levels very efficiently by threshold sweep, where
edges are added to the network in ascending order of their weights. This typical approach in network percolation studies \cite{Newman2010Introduction}
can be even completed without explicitly constructing the network but by only updating a disjoint-sets forest data structure
\cite{Cormen1990Introduction}. As the DAG representation is a static weighted graph these procedures can be used for finding $d_{c_{weak},s_D}$ (and quantities derived from it) for
very large temporal networks and for a range of $\delta t$ values: in practise the limit is the number of events that can be stored in computer memory.

For problems related to
reachability in the DAG one can simplify the full network of all $\delta t$ connections by removing loops via transitive
reduction. Removing all loops in large networks is an expensive procedure, but removing local loops, for example, around
a single node in the network construction process described above (\ref{sec:DAGalg}) is fast. Note that in the special
case of the networks are undirected and no durations are present this local procedure gives the same result
as the one described recently and independently in Ref. \cite{Mellor2017The}.

There are several other ways of making computing various quantities faster that are apparent when the temporal network is represented as a DAG.
A naive way of finding the event that leads to largest number of other events $\rho_{c_{out},s_D}$ would be to start a search from each event
in the network. In the DAG it is apparent that one can omit this search for nodes which have non-zero in-degree and which belong to weakly-connected
components which are smaller than the maximum size found in previous searches. One can further easily calculate the number of nodes downstream of each node
in a weakly-connected components and compare that number to the current maximum size to limit the starting nodes for the searches even further.

\section{Measuring component sizes and order parameters}
\label{sec:DAGcompsize}

\section{Detailed data description}

\begin{table}[h!]
\centering
\caption{Summary of details about the utilised datasets}
\label{table:data}
\begin{tabular}{l|llll}
Dataset & $|E|$ & $|V|$ & duration & resolution \\ \hline  \hline
Mobile calls & $324,528,907$ & $5,193,086$ & $120$ days & $1$ sec \\ \hline
Sexual interactions & $42,409$ & $16,726$ & $2,231$ days & $1$ day \\ \hline
Air transportation & $56,112$ & $279$ & $10$ days $3$ hours & $1$ min \\ \hline
\end{tabular}
\end{table}

We utilised three datasets in this study, each recording a temporal network as a sequence of events. They were:

\begin{itemize}
\item Mobile call interaction network~\cite{Karsai2011} recording 324,528,907 millions of temporal interactions over 120 days of 5,193,086 millions of customers of a single provider in an undisclosed European country. 
\item Sexual-interaction network~\cite{rocha2011simulated} recorded in Brazil with the involvement of 16, 726 sex workers and clients who interacted 42,409 times over 2, 231 days.
\item Air-transportation network containing the time, origin, destination and duration of 180,192 flights between 279 airports in the United States~\cite{USAirline} over 10.3 days. \end{itemize}
Details of each dataset are also summarised in Table~\ref{table:data}. We chose these three empirical networks as they record rather different types of interactions (resp. communication, social, and transportation) and in turn are important to disseminate different types of dynamical processes (resp. the diffusion of information, epidemics, and passengers)

In data two events with no duration (i.e., zero duration) can occur in exactly the same time due to limitation in temporal resolution or due to other reason. This type of events are rare in our data sets, but they can induce loops in the network of events. In order to retain the acyclic property of these graphs we kept only a randomly selected event in case of simultaneous events of the same node.


\begin{thebibliography}{10}

\bibitem{barrat2008dynamical}
Alain Barrat, Marc Barthelemy, and Alessandro Vespignani.
\newblock {\em Dynamical processes on complex networks}, volume 574.
\newblock Cambridge University Press Cambridge, 2008.

\bibitem{pastorsatorras2015}
Romualdo Pastor-Satorras, Claudio Castellano, Piet Van~Mieghem, and Alessandro
  Vespignani.
\newblock Epidemic processes in complex networks.
\newblock {\em Rev. Mod. Phys.}, 87:925, 2015.

\bibitem{saramaki_holme_2012}
Petter Holme and Jari Saram{\"a}ki.
\newblock Temporal networks.
\newblock {\em Phys. Rep.}, 519(3):97--125, 2012.

\bibitem{Holme2015}
Petter Holme.
\newblock Modern temporal network theory: a colloquium.
\newblock {\em Eur. Phys. J. B}, 88(9):234, 2015.

\bibitem{masudalambiotte2016}
Naoki Masuda and Renaud Lambiotte.
\newblock {\em A Guide to Temporal Networks}.
\newblock World Scientific, 2016.
\newblock Series on Complexity Science Vol. 4.

\bibitem{Kempe2002}
D.~Kempe, J.~Kleinberg, and A.~Kumar.
\newblock Connectivity and inference problems for temporal networks.
\newblock {\em Journal of Computer and System Sciences}, 64:820, 2002.

\bibitem{Moody2002}
J.~Moody.
\newblock The importance of relationship timing for diffusion.
\newblock {\em Social Forces}, 81:25--56, 2002.

\bibitem{holme2005network}
Petter Holme.
\newblock Network reachability of real-world contact sequences.
\newblock {\em Phys. Rev. E}, 71(4):046119, 2005.

\bibitem{pan2011path}
Raj~Kumar Pan and Jari Saram{\"a}ki.
\newblock Path lengths, correlations, and centrality in temporal networks.
\newblock {\em Phys. Rev. E}, 84(1):016105, 2011.

\bibitem{scholtes2014causality}
Ingo Scholtes, Nicolas Wider, Ren{\'e} Pfitzner, Antonios Garas, Claudio~J
  Tessone, and Frank Schweitzer.
\newblock Causality-driven slow-down and speed-up of diffusion in non-markovian
  temporal networks.
\newblock {\em Nat. Commun.}, 5, 2014.

\bibitem{Iribarren2009}
J.L. Iribarren and E.~Moro.
\newblock Impact of human activity patterns on the dynamics of information
  diffusion.
\newblock {\em Phys. Rev. Lett.}, 103:038702, 2009.

\bibitem{Karsai2011}
M.~Karsai, M.~Kivel\"a, R.~K. Pan, K.~Kaski, J.~Kert\'esz, A.-L. Barab\'asi,
  and J.~Saram\"aki.
\newblock Small but slow world: How network topology and burstiness slow down
  spreading.
\newblock {\em Phys. Rev. E}, 83:025102, 2011.

\bibitem{kivela2012multiscale}
Mikko Kivel{\"a}, Raj~Kumar Pan, Kimmo Kaski, J{\'a}nos Kert{\'e}sz, Jari
  Saram{\"a}ki, and M{\'a}rton Karsai.
\newblock Multiscale analysis of spreading in a large communication network.
\newblock {\em J. Stat. Mech. Theor. Exp.}, 2012(03):P03005, 2012.

\bibitem{Horvath2014}
D\'avid~X Horv\'ath and J\'anos Kert\'esz.
\newblock Spreading dynamics on networks: the role of burstiness, topology and
  non-stationarity.
\newblock {\em New J. Phys.}, 16(7):073037, 2014.

\bibitem{delvenne2015}
Jean-Charles Delvenne, Renaud Lambiotte, and Luis E.~C. Rocha.
\newblock Diffusion on networked systems is a question of time or structure.
\newblock {\em Nat. Commun.}, 6, 06 2015.

\bibitem{pfitzner2013}
Ren\'e Pfitzner, Ingo Scholtes, Antonios Garas, Claudio~J. Tessone, and Frank
  Schweitzer.
\newblock Betweenness preference: Quantifying correlations in the topological
  dynamics of temporal networks.
\newblock {\em Phys. Rev. Lett.}, 110:198701, 2013.

\bibitem{Backlund2014}
Ville-Pekka Backlund, Jari Saram\"aki, and Raj~Kumar Pan.
\newblock Effects of temporal correlations on cascades: Threshold models on
  temporal networks.
\newblock {\em Phys. Rev. E}, 89:062815, 2014.

\bibitem{Aoki2016}
Takaaki Aoki, Taro Takaguchi, Ryota Kobayashi, and Renaud Lambiotte.
\newblock Input-output relationship in social communications characterized by
  spike train analysis.
\newblock {\em Phys. Rev. E}, 94:042313, 2016.

\bibitem{Miritello2011}
Giovanna Miritello, Esteban Moro, and Rub\'en Lara.
\newblock Dynamical strength of social ties in information spreading.
\newblock {\em Phys. Rev. E}, 83:045102, 2011.

\bibitem{rocha2011simulated}
Luis E~C Rocha, Fredrik Liljeros, and Petter Holme.
\newblock Simulated epidemics in an empirical spatiotemporal network of 50,185
  sexual contacts.
\newblock {\em PLoS computational biology}, 7(3):e1001109, 2011.

\bibitem{Lee2012}
S.~Lee, L.E.C. Rocha, F.~Liljeros, and P.~Holme.
\newblock Exploiting temporal network structures of human interaction to
  effectively immunize populations.
\newblock {\em PLoS One}, 7:e36439+, 2012.

\bibitem{holme2014birth}
Petter Holme and Fredrik Liljeros.
\newblock Birth and death of links control disease spreading in empirical
  contact networks.
\newblock {\em Sci. Rep.}, 4:4999, 2014.

\bibitem{holmemasuda2015}
Petter Holme and Naoki Masuda.
\newblock The basic reproduction number as a predictor for epidemic outbreaks
  in temporal networks.
\newblock {\em PLoS One}, 10, 2015.

\bibitem{valdano2015}
Eugenio Valdano, Luca Ferreri, Chiara Poletto, and Vittoria Colizza.
\newblock Analytical computation of the epidemic threshold on temporal
  networks.
\newblock {\em Phys. Rev. X}, 5:021005, 2015.

\bibitem{Genois2015}
M.~G\'enois, C.L. Vestergaard, C~Cattuto, and A~Barrat.
\newblock Compensating for population sampling in simulations of epidemic
  spread on temporal contact networks.
\newblock {\em Nat. Comm.}, 6:8860, 2015.

\bibitem{HolmePRE2016}
Petter Holme.
\newblock Temporal network structures controlling disease spreading.
\newblock {\em Phys. Rev. E}, 94:022305, 2016.

\bibitem{Speidel2016}
L.~Speidel, K.~Klemm, V.~Egu\'iluz, and N.~Masuda.
\newblock Temporal interactions facilitate endemicity in the
  susceptible-infected-susceptible epidemic model.
\newblock {\em New J. Phys.}, 18:073013, 2016.

\bibitem{DaleyKendall}
D.J. Daley and D.G. Kendall.
\newblock Epidemics and rumours.
\newblock {\em Nature}, 204:1118, 1964.

\bibitem{socdynrev}
Claudio Castellano, Santo Fortunato, and Vittorio Loreto.
\newblock Statistical physics of social dynamics.
\newblock {\em Rev. Mod. Phys.}, 81:591--646, 2009.

\bibitem{TrippBarba20161}
Carolina Tripp-Barba, Cristina Alcaraz, and M\'onica~Aguilar Igartua.
\newblock Special issue on "modeling and performance evaluation of wireless
  ad-hoc networks.
\newblock {\em Ad Hoc Networks}, 52:1 -- 2, 2016.

\bibitem{Nassir201626}
Neema Nassir, Mark Hickman, Ali Malekzadeh, and Elnaz Irannezhad.
\newblock A utility-based travel impedance measure for public transit network
  accessibility.
\newblock {\em Transportation Research Part A: Policy and Practice}, 88:26 --
  39, 2016.

\bibitem{Mellor2017The}
Andrew Mellor.
\newblock {The Temporal Event Graph}, June 2017.

\bibitem{backlund2014Thesis}
Ville-Pekka Backlund.
\newblock Temporal percolation and influential nodes in communication networks.
\newblock M.Sc. thesis, Aalto University, 2014.

\bibitem{ayala2015}
Daniella Ayala.
\newblock Temporal percolation in the erd\"os-r\'enyi model and the effect of
  burstiness.
\newblock M.Sc. thesis, Univ. Oxford, 2015.

\bibitem{Hinrichsen2000Non}
Haye Hinrichsen.
\newblock Non-equilibrium critical phenomena and phase transitions into
  absorbing states.
\newblock {\em Advances in Physics}, 49(7):815--958, 2000.

\bibitem{Karsai2006Nonequilibrium}
M\'arton Karsai, R\'obert Juh\'asz, and Ferenc Igl\'oi.
\newblock Nonequilibrium phase transitions and finite-size scaling in weighted
  scale-free networks.
\newblock {\em Phys. Rev. E}, 73:036116, Mar 2006.

\bibitem{USAirline}
{Bureau of Transportation Statistics}, 2017.

\bibitem{Kovanen2011}
Lauri Kovanen, M\'arton Karsai, Kimmo Kaski, J\'anos Kert\'esz, and Jari
  Saram\"aki.
\newblock Temporal motifs in time-dependent networks.
\newblock {\em J. Stat. Mech. Theor. Exp.}, 2011(11):P11005, 2011.

\bibitem{Holme2012Temporal}
Petter Holme and Jari Saram\"{a}ki.
\newblock {Temporal networks}.
\newblock {\em Physics Reports}, 519(3):97--125, October 2012.

\bibitem{Newman2010Introduction}
Mark Newman.
\newblock {\em {Networks: An Introduction}}.
\newblock Oxford University Press, 1 edition, May 2010.

\bibitem{Cormen1990Introduction}
Thomas~H. Cormen, Charles~E. Leiserson, Ronald~L. Rivest, and Clifford Stein.
\newblock {\em {Introduction to Algorithms, Second Edition}}.
\newblock The MIT Press, 2nd edition, September 2001.

\end{thebibliography}
\end{document}